\documentstyle[epsf]{elsart}

\begin{document}

hep-ph/9506346
\hfill THU-95/16

\begin{center}
{\LARGE \bf Bound state solutions of scalar \\
\vspace{0.3cm}QED$_{2+1}$ for
zero photon mass}

\vspace{0.8truecm}

{\large \bf \underline{Taco Nieuwenhuis}\footnote{E-mail adress: \tt
nieuwenh@fys.ruu.nl} and J.\ A.\ Tjon\footnote{E-mail adress: \tt
tjon@fys.ruu.nl}}

{\small \sl
 Institute for Theoretical Physics, University of Utrecht,
Princetonplein 5,\\
 P.O.\ Box 80.006, 3508 TA Utrecht, the Netherlands.}
\vspace{0.8cm}

{\bf Abstract}
\end{center}
The Feynman-Schwinger representation is used to study
the behavior of solutions of scalar QED in (2+1) dimensions.
The limit of zero photon mass is seen to be
smooth.
The Bethe-Salpeter equation in the ladder
approximation also exhibits this property.
They clearly deviate from
the behavior in the nonrelativistic limit.
In a variational analysis we show that
this difference can be attributed to
retardation effects of relativistic origin.

\vspace{0.8cm}
{\small
PACS numbers: {\sl 11.15.Tk, 11.10.St, 03.65.Ge}
\vspace{-0.0cm}
\begin{tabbing}
Keywords: \= {\sl gauge invariance, nonperturbative scalar
QED$_{2+1}$, bound states,} \\
\> {\sl nonrelativistic limit, Feynman-Schwinger representation,
Bethe-Salpeter} \\
\> {\sl equation, logarithmic confinement.}
\end{tabbing}}

\vspace{0.8cm}
Accepted for publication in {\sl Physics Letters B}

\newpage

An important issue in the description of relativistic
composite systems is to find a reliable and practical
formalism that is consistent with known limiting cases
and the underlying symmetries of the system at hand.
The Feynman-Schwinger representation (FSR)
[1--7]
offers a nonperturbative description which is consistent
with both gauge and Lorentz invariance. Furthermore, the
formalism was shown to satisfy the correct static and
nonrelativistic limits. In this formulation all ladder
and crossed graphs are summed up, while the inclusion
of valence particle loops, self energies and vertex corrections
is feasible as well. Moreover, the FSR is well suited for the
study of nonabelian confining theories such as QCD, since
vacuum condensates can easily be accounted for via the
cumulant expansion [3--5]. These considerations indicate that
the FSR is an appealing alternative to the celebrated
Bethe-Salpeter equation (BSE).

In this letter we apply the FSR to (2+1) dimensional scalar
QED (sQED$_{2+1}$) in the two-body sector. Our interest in
this theory is twofold. First we consider it to be an excellent
playground to study the FSR in the equal mass case, since the
theory is UV finite in the two-body sector in contrast with
the situation in (3+1) dimensions. Secondly, it has been proven
that compactified quenched spinor QED$_{2+1}$ is linear
confining for all values of the charge \cite{qed,fiebig}.
Since the spin of the valence particles is
usually not considered to be essential for the mechanism
behind confinement, one may hope to learn something about the
confinement mechanism in this abelian case. In this work however,
we will focus on the restoration of gauge invariance.

In order to
avoid IR problems we introduce a photon mass $\mu$ in the theory
and study the limit $\mu \rightarrow 0$. In this limit
we may compare the FSR-results with those of the BSE in the ladder
approximation and the Schr\"odinger equation. The $\mu$ dependence
can readily be found in the latter case.
Taking the one meson exchange  contribution
as driving force between 2 charges with mass $m$,
we have in 2 spatial dimensions:
\begin{equation}
\left( -\frac{1}{m}\Delta -\frac{e^2}{2\pi }K_0(\mu r)\right)\psi
(r)= -E_B\psi (r)
\label{schr}
\end{equation}
As $\mu \rightarrow 0$ the modified Bessel function $K_0(\mu r)$
behaves as $-{\rm log}(\mu r)$, which leads to the well known
result of a logarithmic confining
Coulomb potential in 2 spatial dimensions
after an (infinite) energy shift. This interaction
causes $E_B$ to
diverge as $-\mbox{$\frac{e^2}{2\pi}$} {\rm log}(\mu/e\sqrt{m})$,
at least for the low lying states.

Let us now turn to the sQED$_{2+1}$ case. We consider 2 scalar
particles with mass $m$, minimally coupled to the massive photon
field $A_\mu$. The Euclidean action for this theory is:
\begin{equation}
S =\int{\rm d}^3x\left[ \left|(\partial_\mu -{\rm i} e A_\mu)\phi
\right|^2 + m^2|\phi|^2 + \mbox{$\frac{1}{4}$}F_{\mu\nu }^2+
\mbox{$\frac{1}{2}$}\mu^2A_\mu^2
+\mbox{$\frac{1}{2}$}\xi^{-1}(\partial_\mu A_\mu)^2\right]
\label{action}
\end{equation}
The parameter $\xi$ takes care of the gauge fixing when we restore
gauge invariance by taking the limit $\mu \rightarrow 0$.

The object under study is the gauge invariant 4-point function of
the theory, defined as the transition matrix element between the
initial state $\Psi_{\rm i} (x,\bar{x})=\phi^{\dagger}(x)
P(x,\bar{x})\phi (\bar{x})$ and the final state $\Psi_{\rm f}
(y,\bar{y})=\phi(y)P(y,\bar{y})\phi^{\dagger} (\bar{y})$:
\begin{equation}
G(x,\bar{x},y,\bar{y})=\int {\cal D} \phi\;  {\cal D} A_\mu \;
\Psi_{\rm f} (y,\bar{y})\; \Psi_{\rm i}(x,\bar{x})\;
{\rm e}^{-S}
\label{gdef}
\end{equation}
The wavefunctions $\Psi_{\rm i}$ and $\Psi_{\rm f}$ are defined
in a gauge invariant fashion by means of the parallel transporter
$P(x,y)\equiv \exp [-{\rm i} e \int_x^y {\rm d} z_\mu A_\mu (z)]$.
Next the valence field $\phi$ is integrated out and the resulting
determinant is set to unity. The latter amounts to
neglecting all $\phi$-loops. It was shown
in \cite{fsr} that the resulting `quenched' Greens' function can be
written in the following form:
\begin{equation}
G(x,\bar{x},y,\bar{y})=\int_0^\infty{\rm d} s\int_0^\infty
{\rm d}\bar{s} \int ( {\cal D} z)_{xy}
( {\cal D} \bar{z})_{\bar{x}\bar{y}} \;{\rm e}^{-K\left[\{z\},s
\right]- K\left[\{\bar{z}\},\bar{s}\right]}\left\langle
W_{\{z,\bar{z}\}}\right\rangle
\label{gfsr}
\end{equation}
where the path integral $\int( {\cal D} z)_{xy}$ is a
quantummechanical one, subject to the condition of fixed endpoints:
$z(0)=x$ and $z(1)=y$. The functional $K$ is the action of a free
relativistic point particle with fixed eigentime $s$ \cite{poly}:
\begin{equation}
K[\{z\},s] = m^2s+\frac{1}{4s}\int_0^1{\rm d}\tau
\;\dot{z}_\mu^2(\tau )
\end{equation}
The Wilson-loop $\left\langle W_{\{z,\bar{z}\}}\right\rangle$
can be computed exactly in the $U(1)$-case considered here:
\begin{eqnarray}
\left\langle W_{\{z,\bar{z}\}}\right\rangle & = & \left\langle
\exp\left[-{\rm i} e \oint_C {\rm d} w_\mu A_\mu (w)\right]
\right\rangle_{A_\mu}
\label{wdef}\\
& = & \exp\left[ -\mbox{$\frac{1}{2}$}e^2 \oint_C {\rm d} v_\mu
\oint_C {\rm d} w_\nu \Delta_{\mu\nu } (v-w)\right]
\label{wu1}
\end{eqnarray}
$\left\langle W_{\{ z,\bar{z}\} }\right\rangle$ is known to be an order
parameter of the deconfinement transition in pure gauge theories
\cite{lgt1,poly}. The variables $v$ and $w$ in (\ref{wdef}) and
(\ref{wu1}) both run along the whole closed
contour $C$ formed by the lines $x\bar{x}$ and $y\bar{y}$ and the
paths $\{z\}$ and $\{\bar{z}\}$.
Any longitudinal
term vanishes identically in (\ref{wu1})\footnote{In order to see this
we Fourier transform to momentum space and concentrate on the
longitudinal contributions:
\begin{eqnarray*}
\lefteqn{\int {\rm d}\tau\!\int{\rm d}\sigma \;\dot{w}_\mu (\tau )\;
\dot{v}_\nu (\sigma )\; \Delta_{\mu\nu }(w-v) = } \\
& & \hspace{3cm} \int \!\frac{{\rm d}^dq}{
(2\pi)^d} \int{\rm d}\tau\!\int{\rm d}\sigma \;{\rm e}^{{\rm i} q
\cdot (w-v)} \;\partial_\tau (q\cdot w(\tau ))
\;\partial_\sigma (q\cdot v(\sigma ))\; f(q)
\end{eqnarray*}
which vanishes identically for every closed contour. Notice that
this is the case for finite $\mu$ as well.}. The nonvanishing
transverse component of $\Delta_{\mu\nu }$ is given by:
$\Delta_{\mu\nu }(r)=\delta_{\mu\nu }{\rm e}^{-\mu r}/4\pi  r \equiv
\delta_{\mu\nu }\Delta (r)$.
It can be shown that since there is no ordering of $v$ and $w$ present
in (\ref{wu1}), it sums up all ladder, crossed ladder, self energy and
vertex correction graphs with the appropriate
weights.
Comparing the FSR  to
quenched lattice gauge calculations,
the calculation of (\ref{gfsr})
can obviously be  carried out in a
numerically more accurate way since the occurring
path integrals are quantummechanical ones.

Here we wish to concentrate
on the ladder and crossed graphs only and we therefore restrict $w$
and $v$ to opposite sides of the contour.
The bound state spectrum of the thus obtained
expression is studied by considering
the asymptotic spectral decomposition of $G$:
\begin{equation}
G(T) = \lim_{T\rightarrow\infty}\sum_n c_n \exp (-E_n T) \hspace{1.5cm}
(T\equiv \mbox{$\frac{1}{2}$}(y_3+\bar{y}_3-x_3-\bar{x}_3))
\label{spec}
\end{equation}
and its logarithmic derivative $L(T)=-\left[\frac{{\rm d}}{{\rm
d}T}G(T)\right] /G(T)$.
According to (\ref{spec}), $L(T)$ is expected to approach the mass of the
ground state $E_0$ as $T\rightarrow\infty$.
In practice we approximated the path integral (\ref{gfsr})
by a finite product of $N-1$ integrals:
\begin{equation}
\int ( {\cal D} z)_{xy} \longrightarrow \left(\frac{N}{4\pi
s}\right)^{3N/2} \;\prod_{i=1}^{N-1}
\int {\rm d}^3z_i
\label{discr}
\end{equation}
with the constraints $z_0=x$ and $z_N=y$. The normalization factor in
(\ref{discr}) is {\em not} irrelevant since it contains $s$ which is
integrated over. The functionals $K$ and $V\equiv\log\left\langle
W\right\rangle$ are discretized as follows:
\begin{eqnarray}
K[\{ z\} ,s] &\longrightarrow & m^2s+\frac{N}{4s}\sum_{i=1}^N \left(
z_i - z_{i-1}\right)^2
\label{kdiscr}\\
& & \nonumber \\
& & \nonumber \\
V[\{ z,\bar{z}\}]&\longrightarrow &\nonumber
\end{eqnarray}
\begin{equation}
\hspace{1cm}e^2 \sum_{i,j=1}^N \left( z_i-z_{i-1}\right)\cdot\left(
\bar{z}_j-\bar{z}_{j-1}\right)\Delta \left( \mbox{$\frac{1}{2}$}\left(
z_i+z_{i-1} -\bar{z}_j-\bar{z}_{j-1}\right)\right)
\label{wdiscr}
\end{equation}
Here we adopted the Weyl ordering prescription for noncommuting
operators, leading to the midpoint discretization in (\ref{wdiscr}).
Our calculations were checked on the convergence
with respect to the value of $N$
and usually $N=30$ was found to be sufficiently
large to detect no $N$-dependence within statistical errors. We projected
$G$ on the eigenstates of total and angular momentum $\left|{\bf P},m
\right\rangle$ but due to the nonlocality in (\ref{wdiscr}) this did
not imply that the degrees of freedom associated with the symmetries
(${\bf P}$, $L^2$) could be integrated out. The logarithmic derivative
$L(T)$ can be symbolically written as a statistical
average (here we denote the set of integration variables $(s$,
$\bar{s}$, $\{ z_i\}$, $\{ \bar{z}_i\})$ shorthandedly
as $Z$):
\begin{equation}
L=\frac{\int\! {\cal D} Z \left( K'[Z] - V'[Z]\right) {\rm e}^{-K[Z]+
V[Z]}}{ \int\! {\cal D} Z  \;{\rm e}^{-K[Z]+V[Z]}}
\label{stat}
\end{equation}
where the prime stands for analytical differentiation with respect to
$T$.
We performed Metropolis Monte Carlo calculations of this
object by averaging $K'[Z]-V'[Z]$ over an ensemble generated by
$\exp \left( -K[Z]+V[Z]\right)$.
\begin{figure}
\epsfxsize=12cm
\epsfysize=9.5cm
\epsffile{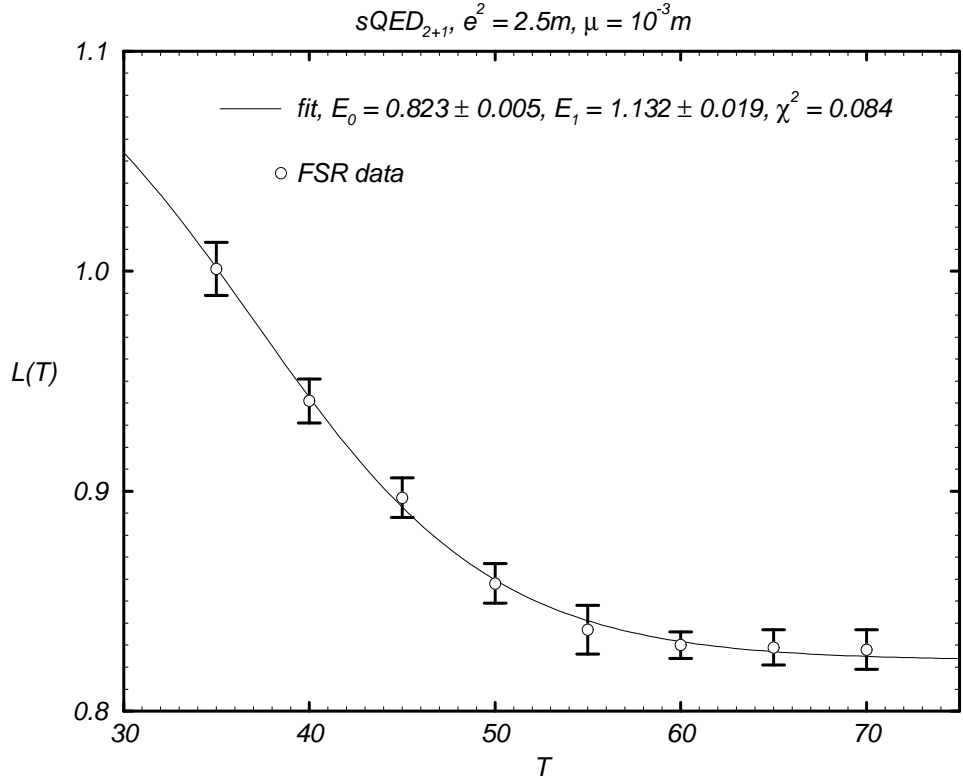}
\caption{\sl Logarithmic derivative $L(T)$ as a function
of the time $T$ for $e^2=2.5m$ and $\mu = 10^{-3}m$. The
solid line is a fit of the data to the form (8).\newline}
\end{figure}
In Fig.\ 1 we present results of calculations of $L(T)$ for
the case $e^2=2.5m$ and $\mu = 10^{-3}m$. It is clear that $L(T)$
approaches a constant for large $T$ and a fit of the data to the
form (\ref{spec}) allows one to get an accurate result (within 1\%)
for $E_0$ and in principle even a good indication (within 5\%) of
the first excited state. Note that we are able to do calculations
at arbitrary large $T$ since spacetime is {\em not} discretized
in the FSR approach. This is in sharp contrast with lattice calculations
where spacetime is necessarily finite. Besides that, the accuracy of
our results is high as compared to lattice calculations
\cite{lgt1}.
\begin{figure}
\epsfxsize=12cm
\epsfysize=9.5cm
\epsffile{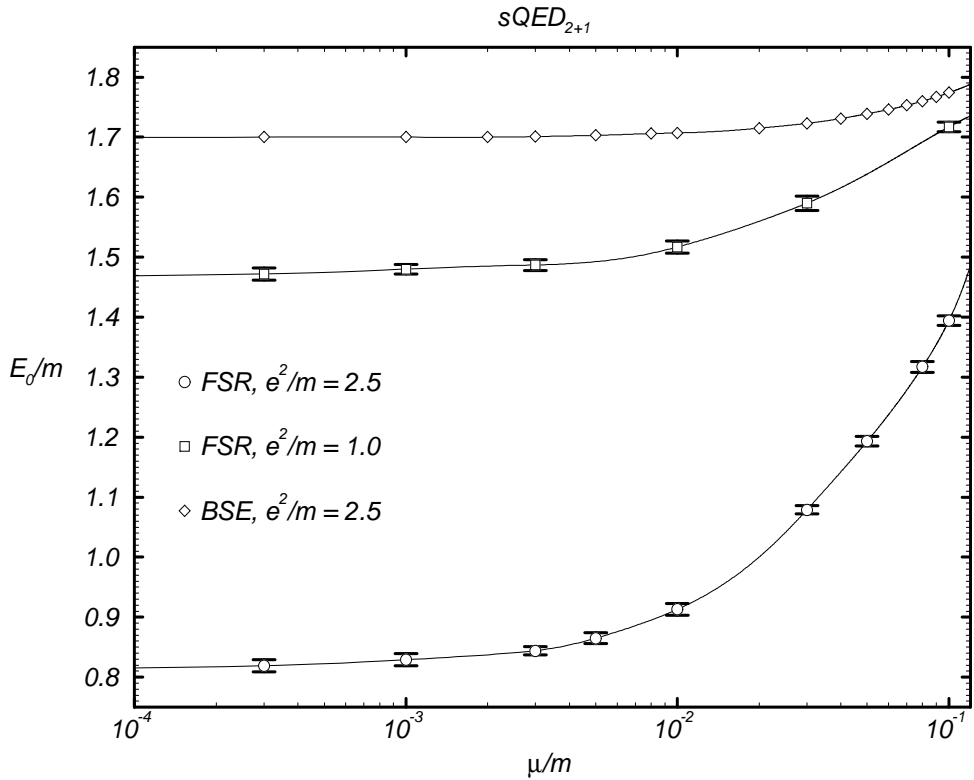}
\caption{\sl Mass $E_0$ of the ground state for
sQED$_{2+1}$ as a function of the photon mass $\mu$. The
lines are just drawn to guide the eye. Any logarithmic
dependence of $E_0$ on $\mu$ would show up in this figure
as a straight nonhorizontal line.
\newline}
\end{figure}
In Fig.\ 2
we display the mass of the ground state obtained with this procedure,
as a function of $\mu$ for two different values of the coupling constant
$e^2$. Clearly the limit $\mu \rightarrow 0$ is a smooth one and we are
able to restore gauge invariance in this way. Hence, no logarithmic
confinement is found in (2+1) dimensions as long as the mass of the
charged particles is kept finite. From these studies we furthermore
find that for each $\mu$ there is a finite critical coupling constant
where $E_0=0$. For $\mu /m = 0.1$ this critical coupling constant was
determined to be $e_{\rm crit}^2/m=6.0\pm 0.5$. At this point the vacuum
becomes degenerate in view of these zero mass two-particle bound states.

To get some insight on the absence of the logarithmic confinement in the
relativistic case we may study the BSE \cite{bse,nak,itzu} in the ladder
approximation for this theory. In the CM system where $P_{\rm tot}
= \left({\bf 0},\sqrt{s}\right)$ it takes on the following form:
\begin{equation}
\Phi (p) = \frac{e^2}{(2\pi  )^3} \int{\rm d}^3q \; V(p,q) \; G_0 (q) \;
\Phi (q)
\label{bse}
\end{equation}
with
\begin{equation}
V(p,q) = \frac{s+(p+q)^2}{(p-q)^2+\mu^2} \hspace{0.7cm}{\rm and}
\hspace{0.7cm} G_0(q) = \frac{1}{(m^2-\mbox{$\frac{1}{4}$}s+q^2)^2
+sq_3^2}
\label{vg}
\end{equation}
After Wick rotation \cite{wrot} the momenta in (\ref{bse}) and (\ref{vg})
can be taken Euclidean.
The solutions to (\ref{bse}) were constructed by expanding $\Phi$, $V$
and $G_0$ on the basis of $O(3)$ spherical harmonics $Y_{lm}(\Omega_3)$
and truncating the resulting infinite set of coupled one-dimensional
integral equations at a certain $l_{\rm max}$. Truncation at $l_{\rm max}
=2$ was sufficient to get very accurate results. Further details of this
procedure will be published elsewhere. In Fig.\ 2 the result
of the calculations for $e^2 = 2.5m$ are indicated by the
diamonds. As compared to the FSR result, the
BSE solutions show much less binding.
This is a general feature of
all our calculations and it is particularly striking for strong couplings.
We are led to conclude that the crossed ladder diagrams give a very
significant contribution to the binding energy of the system considered
here. The limit $\mu \rightarrow 0$ is seen
to be smooth as well and we can safely remove the IR cut-off this way.

One may address the question what mechanism is responsible for the
smoothness of the limit $\mu\rightarrow 0$ in the relativistic
calculations. It is generally known that both the BSE and the FSR
have the correct nonrelativistic limit and besides that, it is rather
remarkable that such a global feature as this logarithmic divergence
does not persist when one considers a relativistic theory.
These considerations raise the question how the nonrelativistic limit is
being reached in this case. For this purpose we carried out a variational
analysis of the BSE for a $g\phi^3_{2+1}$-theory. It is convenient to
introduce the nonrelativistic coupling constant $\lambda = g^2/4 m^2$
which enters the corresponding Schr\"odinger equation. We verified that
relativistically this case behaves smooth as well for $\mu\rightarrow 0$,
while it becomes identical to sQED$_{2+1}$ in the nonrelativistic limit.
Our trial function used:
\begin{equation}
\phi_{\rm trial}({\bf r},t)=\exp\left[-\mbox{$\frac{1}{2}$}\alpha^2r^2
\right] \exp\left[-\mbox{$\frac{1}{2}$}\beta^2t^2\right]
\label{try}
\end{equation}
with variational parameters $\alpha$ and $\beta$,
is particularly suited for evaluating the various matrix elements
analytically.  Nonrelativistically we expect $\alpha\sim {\cal O}
\left(\sqrt{ E_B/m}\right)\gg\beta\sim {\cal O} \left(E_B/m\right)$.
Applying the Rayleigh-Ritz variational principle to
the energy functional $s(\alpha ,\beta )$ indeed yields this property
up to multiplicative logarithmic corrections.
The nonrelativistic limit is obtained by letting $m/\mu \rightarrow
\infty$ {\em and} at the same time $\lambda/\mu\rightarrow 0$, while
keeping their product $\zeta =m\lambda/\mu^2$ constant. In this region
the wavefunction becomes independent of the relative time $t$ and as a
result the Schr\"odinger predictions are obtained.  We find in
particular that for
\begin{equation}
\frac{\lambda}{m}\log^2 (m/\lambda ) \ll \zeta^{-1} \ll 1
\label{nrc}
\end{equation}
the binding energy
$E_B=\left( m^2-\mbox{$\frac{1}{4}$} s\right)/m$ goes as $\sim \mbox{
$\frac{1}{4\pi }$}\lambda\log \zeta$. Hence in this region the logarithmic
divergence of the Schr\"odinger analysis is recovered.  We see however
that $E_B$ {\em only} exhibits this logarithmic dependence on the photon
mass $\mu$ as long as $\zeta$ is kept at a {\em fixed} value and $m
\rightarrow\infty$ so that the condition (\ref{nrc}) is satisfied.
Taking on the other hand $m$ large but fixed and letting $\mu\rightarrow 0$,
we effectively put $\zeta\rightarrow \infty$, thereby invalidating
(\ref{nrc}).  It can be shown that in this limit $E_B$
approaches a large but finite constant $\sim\lambda\log (m/\lambda )$
independent of $\mu$. This shows that the nonrelativistic limit is not
uniform. For decreasing $\mu$ there is a crossover point $\mu_0$ ($\mu_0
\sim \lambda\log (m/\lambda )$ as can be inferred from (\ref{nrc}))
where the relative time dependence in the
wavefunction starts to play a role.
A dynamical screening mass $\mu_0$
is effectively generated, which is related to
the nonvanishing of the relative time parameter $\beta$ in
(\ref{try}).
It is interesting to note that although $\mu_0$ is vanishingly small
on the scale of $m$, it is essentially of relativistic origin and no
trace of it is left in the Schr\"odinger equation.

It is known that accurate results can be obtained variationally,
even with rather simple trial functions \cite{var}.
Over a wide range of coupling constants we indeed find that the above
variational calculations yield within a few percent the exact BSE results.
In conclusion, we have shown
that the FSR is a very suitable nonperturbative
method to
extract in a reliable way the bound state energy.
Furthermore, the IR cut-off in sQED$_{2+1}$
can safely be removed.
No logarithmic confinement as $\mu\rightarrow 0$
is found in (2+1) dimension for a finite mass
of the charged particles.
This is due to the relative time effects
occurring in a relativistic description.

\vspace{0.5cm}
\noindent
{\bf Acknowledgement}

It is a pleasure to thank Yu.\ A.\ Simonov for many illuminating
discussions concerning the FSR.

\end{document}